\documentclass[nofootinbib, preprintnumbers, showpacs, twocolumn]{revtex4-1}
\usepackage{amsmath, amssymb}
\usepackage[colorlinks = true]{hyperref}
\usepackage{graphicx}

\begin{document}
\preprint{KOBE-COSMO-16-10}
\title{Detecting ultralight axion dark matter wind with laser interferometers}
\author{Arata Aoki}
\email{arata.aoki@stu.kobe-u.ac.jp}
\author{Jiro Soda}
\email{jiro@phys.sci.kobe-u.ac.jp}
\affiliation{Department of Physics, Kobe University, Kobe 657-8501, Japan}
\date{\today}

\begin{abstract}
The ultralight axion with mass around $10^{-22}$~eV is known as a candidate of dark matter.
A peculiar feature of the ultralight axion is oscillating pressure in time, which produces oscillation of gravitational potentials.
Since the solar system moves through the dark matter halo at the velocity of about $v \sim 300 \, \text{km} / \text{s} = 10^{-3}$, there exists axion wind, which looks like scalar gravitational waves for us.
Hence, there is a chance to detect ultralight axion dark matter with a wide mass range by using laser interferometer detectors.
We calculate the detector signal induced by the oscillating pressure of the ultralight axion field, which would be detected by future laser interferometer experiments.
We also argue that the detector signal can be enhanced due to the resonance in modified gravity theory explaining the dark energy.
\end{abstract}

\maketitle

\section{Introduction}
It is known that the cold dark matter (CDM) with a cosmological constant ($\Lambda$CDM model) is currently the most successful cosmological model.
The most promising candidate of the CDM is supersymmetric particles, the so-called neutralino.
While CDM works quite well especially on large scales, there exists a problem on small scales.
In fact, this model predicts a cusp of dark matter halo profile and overabundance of dwarf galaxies, which are not consistent with observations.
Moreover, the LHC has not reported any signature of supersymmetry.
Given this situation, it is worth investigating another possibility, namely axion dark matter.

The axion, a pseudo-Nambu-Goldstone boson, was originally introduced by Peccei and Quinn to resolve the strong CP problem of QCD~\cite{77:Peccei, 77:Peccei-2, 78:Weinberg, 78:Wilczek}.
Nowadays, however, it is known that the string theory also predicts such (pseudo-)scalar fields with a wide range of mass scales~\cite{06:Svrcek, 10:Arvanitaki}.
Remarkably, the axion interacting very weakly with standard model particles is regarded as a candidate of dark matter.
In particular, the ultralight axion with mass around $10^{-22}$~eV can resolve the cusp problem on subgalactic scales~\cite{00:Hu}.\footnote{
	Actually, there is a long history for the scalar dark matter.
	Some early works are Refs.~\onlinecite{83:Baldeschi, 94:Sin, 96:Lee}.
	For complete references see Ref.~\onlinecite{16:Lee}.
}
For this reason, the ultralight axion has recently attracted much attentions.

A peculiarity of the ultralight axion is the oscillating pressure in time with angular frequency at the twice of the axion mass, $\omega = 2m$.
Therefore, in order to identify the axion dark matter, we should detect this oscillating pressure.
Khmelnitsky and Rubakov have recently pointed out that the effect of oscillating pressure might be detected with pulsar timing array experiments~\cite{14:Khmelnitsky}.
Indeed, the oscillating pressure induces oscillation of gravitational potentials with frequency in nano-Hz range.
This effect can be observed as a shift of an arrival time of a signal from a pulsar.
The idea is attractive since it depends only on the gravitational interaction, which is the only guaranteed interaction for dark matter.
However, this method is useful for a very narrow range of mass scales of the axion.
Hence, it is desired to invent other detection methods depending only on the gravitational interaction.

It should be noted that the laser interferometers are useful for detecting gravitational waves.
Remarkably, the era of gravitational wave astronomy just started on 14th September 2015, when the two interferometer detectors of the LIGO simultaneously observed a gravitational wave signal~\cite{16:Abbott}.
Now, we can measure tiny fluctuations of the spacetime by using these high precision interferometer detectors.
Since the solar system moves through the dark matter halo at the velocity of about $v \sim 300 \, \text{km} / \text{s} = 10^{-3}$, the axion wind looks like scalar gravitational waves for us.
Therefore, we can utilize the interferometer detectors for observing the ultralight axion wind.

\section{Axion oscillation}
We consider the situation where the dark matter halo is composed out of the ultralight axions.
We should note that the mass of the axion cannot be lighter than $10^{-23}$~eV to be consistent with structure formation~\cite{14:Khmelnitsky}.
Since the occupation number of the axion in the dark matter halo is huge, we can treat it as a classical scalar field.
The axion field satisfies the Klein-Gordon equation in the flat space-time at the leading order, and the solution is given by superposition of plane waves with different wavenumbers and frequencies.
The wavenumber has a certain cutoff, $k_{\text{max}} \sim mv$, due to the uncertainty principle.
This corresponds to the inverse of the de Broglie wavelength of the axion.
Since the typical velocity in the galaxy is $v \sim 10^{-3}$, we can assume that the axion field oscillates monochromatically with the angular frequency corresponding to its mass.
Under these assumptions, we can write the axion field as
\begin{equation}
	\phi = \phi_{0}\cos(mt) \, ,
\end{equation}
where $\phi_{0}$ is a constant independent of spacetime.
The energy density $\rho$ and the pressure $p$ of the axion field are given by
\begin{equation}
	\rho = \frac{1}{2}m^{2}\phi_{0}^{2} \, , \quad p = -\rho\cos(\omega t) \ ,
\end{equation}
where $\omega \equiv 2m$.
A typical energy density of the dark matter halo is $\rho \sim 0.3 \, \text{GeV} / \text{cm}^{3}$.
The energy-momentum tensor of the axion field is then given by
\begin{equation}
	T_{\mu\nu} = \begin{pmatrix} \rho & 0 \\ 0 & -\rho\cos(\omega t) \delta_{{ij}} \end{pmatrix} \, .
\end{equation}

The period of the oscillation corresponds to about one year, and this time scale is much shorter than the cosmological time scale, i.e. $H_0^{-1} \sim 10^{10}$ years.
Hence, after averaging the oscillating pressure over the cosmological time scale, the axion behaves as pressureless dust on cosmological scales.

\section{Gravitational potentials in detector frame}
In this section,we solve the Einstein equation in a detector reference frame using the Newtonian gauge.
Note that the solar system moves through the dark matter halo at the velocity of about $v \sim 300 \, \text{km} / \text{s} = 10^{-3}$.

In order to calculate the signal on the detector, we should evaluate the energy-momentum tensor in the reference frame of a detector.
The Lorentz-boost transformation in the direction $v^i$ is given by
\begin{align}
	t' &= \gamma(t + \vec{v} \cdot \vec{x}) \, , \\
	x^{i\prime} &=  x^{i} +\frac{\gamma - 1}{v^{2}} x^{k}v^{k}v^{i} + \gamma v^{i}t  \, ,
\end{align}
where $x^{\mu\prime}$ is the coordinate system attached to the detector and $\vec{v}$ is the relative velocity of the detector to the dark matter halo, $\gamma \equiv 1 / \sqrt{1 - v^{2}}$ is the Lorentz factor.
Hence, the energy-momentum tensor in the detector frame is given by
\begin{align}
	T_{00} &= \rho\gamma^{2}[1 - v^{2}\cos(\omega t')] \, , \\
	T_{0i} &= \rho\gamma^{2}v_{i}[1 - \cos(\omega t')] \, , \\
	T_{ij} &= -\rho\cos(\omega t')\delta_{ij} + \rho\gamma^{2}v_{i}v_{j}[1 - \cos(\omega t')] \, ,
\end{align}
where $t' = \gamma(t + \vec{v} \cdot \vec{x})$.

On the scale of the dark matter halo, the expansion of the universe is completely negligible and gravitational potentials can still be treated as perturbations.
Thus, we use the Newtonian gauge for the metric:
\begin{equation}
	g_{\mu\nu} =	\begin{pmatrix}
					-1 - 2\Psi & 0 \\ 0 & (1 - 2\Phi)\delta_{ij}
				\end{pmatrix} + \delta\tilde{g}_{\mu\nu} \, ,
\end{equation}
where $\delta\tilde{g}_{\mu\nu}$ is a constant tensor introduced for the consistency of the Einstein equation, which is produced by the constant velocity $\vec{v}$.
We can calculate the Einstein tensor at the linear order as
\begin{align}
	G_{00} &= 2\nabla^{2}\Phi \, , \\
	G_{0i} &= 2\dot{\Phi}_{,i} + \tilde{G}_{i} \, , \\
	G_{ij} &= [2\ddot{\Phi} - \nabla^{2}(\Phi - \Psi)]\delta_{ij} + \partial_{i}\partial_{j}(\Phi - \Psi) + \tilde{G}_{ij} \, ,
\end{align}
where $\tilde{G}_{i}$ and $\tilde{G}_{ij}$ are constant parts calculated from $\delta\tilde{g}_{\mu\nu}$.

Let us separate the gravitational potential $\Phi \, (\Psi)$ into the time-independent part $\Phi_{0} \, (\Psi_{0})$ and the time-dependent part $\delta\Phi \, (\delta\Psi)$.
As we will see later, $\delta\Psi$ is the only observable quantity by interferometers.
The time-independent part of the $(0, 0)$ component of the Einstein equation is the Poisson equation:
\begin{equation}
	2\nabla^{2}\Phi_{0} = \rho\gamma^{2} \, ,
\end{equation}
where $\gamma^{2} = 1 + \mathcal{O}(v^{2})$.
The time-dependent part of the $(0, 0)$ component of the Einstein equation gives
\begin{equation}
	\delta\Phi(t, \vec{x}) = \frac{\rho}{8m^{2}}\cos[\omega\gamma(t + \vec{v} \cdot \vec{x})] \, .
\end{equation}
Finally, we obtain $\delta\Psi$ from the time-dependent part of the $(i, j)$ component of the Einstein equation:
\begin{equation}
	\delta\Psi(t, \vec{x}) = -\frac{\rho}{8m^{2}}\cos[\omega\gamma(t + \vec{v} \cdot \vec{x})] \, .
\end{equation}
We can check that the other components of the Einstein equation are also satisfied.

\section{Detector signal induced by axion pressure}
Now, we calculate the detector signal induced by the oscillating pressure of the axion field.
First, we calculate the metric in the synchronous(-like) gauge by a gauge transformation in order to know spatial fluctuations of the metric.
The detector signal is then obtained by contracting the metric with a detector tensor.

The mechanism of an interferometer is simple.
A laser light is sent on a beam-splitter which separates the light, with equal probability amplitudes, into a beam traveling in one arm and a beam traveling in a second orthogonal arm.
At the end of each arm, there are totally reflecting mirrors.
After traveling back and forth, the two beams recombine at the beam-splitter, and a part of the resulting beam goes to a photo-detector.
Therefore, any variation in the length of the arms results in a corresponding variation of the power at the photo-detector.
Indeed, using this interferometer, we can detect gravitational waves.
Here, the idea is to utilize the interferometer detector for detecting axion dark matter.

The interferometer detector is characterized by the detector tensor~\cite{00:Maggiore}
\begin{equation}
	D_{ij} \equiv \frac{1}{2}(\hat{m}_{i}\hat{m}_{j} - \hat{n}_{i}\hat{n}_{j}) \, ,
\end{equation}
where $\hat{m}$ and $\hat{n}$ are the directions of the detector arms.
The detector signal produced by spatial fluctuations of the metric $h_{ij}$ is given by
\begin{equation}
	s = D_{ij}h_{ij} \, .
\end{equation}
In order to use this formula, we should transform all of the time-dependent part of the metric to $h_{ij}$.
Namely, we want to rewrite the time-dependent part of the metric as follows:
\begin{equation}
	\delta g_{\mu\nu} = \begin{pmatrix} 0 & 0 \\ 0 & -2 \, \delta A \, \delta_{ij} + 2 \, \delta B_{,ij} \end{pmatrix} \, .
\end{equation}
Note that we focus only on the time-dependent components, and thus this is not exactly the synchronous gauge.
The signal in this gauge is given by
\begin{equation}
	s = (\hat{m}_{i}\hat{m}_{j} - \hat{n}_{i}\hat{n}_{j})\delta B_{,ij} \, .
\end{equation}
Since $D_{ij}\delta_{ij} = 0$ for any detector tensors, $\delta A$ can not be detected by interferometers.
The gauge transformation gives $\delta A = \delta\Phi$ and
\begin{equation}
	\delta\ddot{B} = -\delta\Psi = \frac{\rho}{8m^{2}}\cos[\omega\gamma(t + \vec{v} \cdot \vec{x})] \, .
\end{equation}
Hence, we obtain
\begin{equation}
	\delta B_{,ij} = \frac{\rho}{8m^{2}}v_{i}v_{j}\cos[\omega\gamma(t + \vec{v} \cdot \vec{x})] \, .
\end{equation}
We note that $\delta B$ has ambiguity of function $f(\vec{x})t + g(\vec{x})$.
However, a constant function $g(\vec{x})$ does not affect the signal and stationarity of the system requires that $f(\vec{x})$ vanishes.
Therefore, the signal from the oscillating pressure of the axion field is given by
\begin{equation}
	s(t) = \alpha \cdot \frac{\rho v^{2}}{8m^{2}}\cos(\omega\gamma t) \, ,
	\label{Eq.22}
\end{equation}
where $\alpha \equiv |(\hat{v} \cdot \hat{m})^{2} - (\hat{v} \cdot \hat{n})^{2}|$ is the geometric factor of $\mathcal{O}(1)$.
We omitted the unimportant phase $\omega\gamma\vec{v} \cdot \vec{x}_{\text{d}}$, where $\vec{x}_{\text{d}}$ is the position of the detector.
The factor $\alpha$ reaches the maximum when $\vec{v}$ is parallel to $\hat{m}$ or $\hat{n}$.
If the angle between two arms of the detector is $\theta$, the maximum value of $\alpha$ is $\alpha_{\text{max}} = 1 - \cos^{2}\theta$.
A typical amplitude of the signal is
\begin{equation}
	\frac{\rho v^{2}}{8m^{2}} = 1.6 \times 10^{-23} \left( \frac{v}{10^{-3}} \right) ^{2} \left( \frac{10^{-22} \, \text{eV}}{m} \right) ^{2} \, ,
	\label{Eq.23}
\end{equation}
where we assumed $\rho = 0.3 \, \text{GeV} / \text{cm}^{3}$.
The corresponding frequency is
\begin{equation}
	f = \frac{\omega\gamma}{2\pi} \simeq 5 \times 10^{-8} \, \text{Hz} \, \left( \frac{m}{10^{-22} \, \text{eV}} \right) \, .
\end{equation}
Apparently, the signal is proportional to $m^{-2}$ and $v^{2}$.
Hence, the lower the mass is, the easier we detect the axion wind.
We notice that while the signal from the ultralight axion dark matter grows as the mass $m$ decreases, the axion with $m < 10^{-23}$~eV would wash out structures on the observed scales and is inconsistent with observations~\cite{14:Khmelnitsky}.

\section{Discussion}
The pulsar timing analysis can probe the axion oscillation in the mass range $10^{-23}$ to $2.3 \times 10^{-23}$~eV.
Here, in principle, we can probe the axion with $m > 2.3 \times 10^{-23}$~eV by using laser interferometers.
For example, for $10^{-22}$~eV axion corresponding to 0.1 $\mu$Hz, we can expect the strain $10^{-23}$.

\begin{figure}
\includegraphics[width = \hsize]{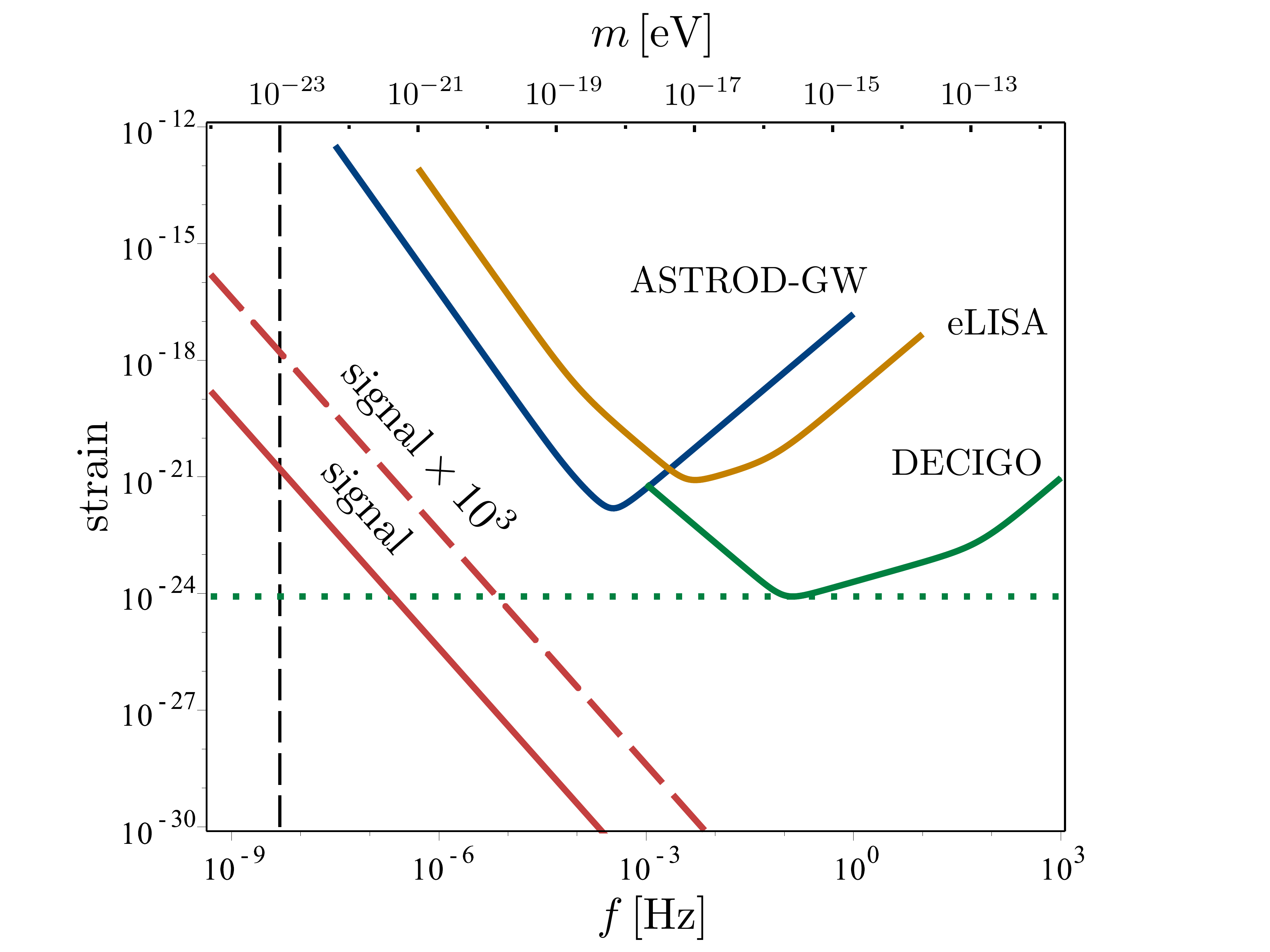}
\caption{
	A typical detector signal (\ref{Eq.23}) and an enhanced signal due to the resonance in a modified gravity with the sensitivity curves of DECIGO, eLISA, and ASTROD-GW.
	Note that the mass of the axion cannot be lighter than $10^{-23}$~eV to be consistent with structure formation.
}
\label{Fig1}
\end{figure}

We depicted a typical detector signal (\ref{Eq.23}) along with the sensitivity curves of DECIGO~\cite{01:Seto}, eLISA~\cite{05:Crowder, 13:Amaro-Seoane}, and ASTROD-GW~\cite{15:Kuroda} in Fig.~\ref{Fig1}.
If we were able to construct a space-based interferometer with strain sensitivity $10^{-24}$, which is similar as that of DECIGO, in the appropriate frequency band, we would be able to observe the axion wind up to $4.0 \times 10^{-22}$~eV.
Though constructing a interferometer with enough sensitivity for detecting the axion wind is beyond current technical capabilities, we believe that it will be realized by future technological innovations.

We should mention that $\vec{v}$, which we have treated as a constant, actually varies in time.
Since the Earth moves around the Sun with a velocity about $30 \, \text{km} / \text{s} = 10^{-4}$, $v$ varies by about $10\%$ in one year.
However, we believe that such an annual modulation in the signal can be extracted as a noise.

Although we have discussed the detectability of axion oscillation based on Einstein's theory, there are many alternative theories which modifies Einstein's theory for resolving the dark energy problem.
Such modified gravity theories typically include dynamical scalar fields, which are known as scalarons, in addition to the gravitational waves.
In our previous paper~\cite{16:Aoki}, we have pointed out that, in the case of $f(R)$ theory, a resonance between the axion and the scalaron can amplify the signal dramatically when their masses are sufficiently close to each other.
Thus, in the case of modified gravity theories, it might be possible to detect the signal in a higher frequency range.
If the signal were amplified by a factor of $\lambda$ due to the resonance, the detectable mass range would expand to $\sqrt{\lambda}$ times.
In Fig.~\ref{Fig1}, we showed the case of $\lambda = 10^{3}$ as an example.

\section{Conclusion}
We have investigated the detectability of the ultralight axion dark matter using gravitational wave laser interferometers.
Specifically, we calculated the detector signal induced by the oscillating pressure of the ultralight axion field.
We obtained Eq. (\ref{Eq.22}), which shows that the signal is proportional to $m^{-2}$ and $v^{2}$, where $m$ is the mass of the axion and $v$ is the relative velocity of the detector relative to the dark matter halo.
A typical signal has a strain of order $10^{-23}$ and a frequency about 0.1~$\mu$Hz if the axion mass is $m = 10^{-22}$~eV.
While a signal is too small to be detected with current technical capabilities, it can in principle be detected by future laser interferometers.

Remarkably, the oscillation of the gravitational potential induced by the axion can be enhanced due to the resonance in modified gravity theories explaining the dark energy in the present universe.
Thus, in the case of modified gravity theories, it would be possible to detect the signal in a more wide mass range by using the  space-based laser interferometers.

There are various detectors to observe gravitational waves in the wide range of frequency from nano-Hz to MHz.
The ground-based experiments, such as Virgo and KAGRA~\cite{12:Somiya}, will begin to operate in the near future.
In addition, space-based interferometers, such as DECIGO and eLISA, are planned for detecting low-frequency gravitational waves.
For even lower frequencies, interferometers with arm length longer than that of eLISA, such as ASTROD-GW, are also studied~\cite{15:Kuroda}.
It is interesting to construct detectors in $\mu$Hz frequency range where we also have a possibility to detect the ultralight axion dark matter.

\acknowledgments
This work was in part supported by MEXT KAKENHI Grant Number 15H05895.

\bibliographystyle{apsrev4-1}
\bibliography{draft}

\end{document}